%% file: ase26.tex
\documentclass[sigconf,screen]{acmart}

\usepackage[nolist]{acronym}
\usepackage{xspace}
\input{acronyms}

\usepackage[english]{babel}
\usepackage{csquotes}

\usepackage{tabularx}

\usepackage{adjustbox}

\usepackage[dvipsnames]{xcolor}

\ifthenelse{0 > -1}{
    \newcommand{\kf}[1]{\textcolor{kit-orange}{[KF] #1}}
    \newcommand{\df}[1]{\textcolor{kit-purple}{[DF] #1}}
}{
    \newcommand{\kf}[1]{\textcolor{kit-orange}{}}
    \newcommand{\df}[1]{\textcolor{kit-purple}{}}
}

\setcopyright{cc}
\setcctype{by}
\copyrightyear{2026}
\acmYear{2026}
\acmDOI{XXXXXXX.XXXXXXX}
\acmConference[ASE '26]{the 41st IEEE/ACM International Conference on Automated Software Engineering}{October 12--16,
  2026}{Munich, Germany}
\acmISBN{978-1-4503-XXXX-X/2026/10}

\usepackage{pifont}

\usepackage{makecell}
\usepackage{booktabs}
\usepackage{enumitem}
\setlist[itemize,enumerate]{leftmargin=*}

\graphicspath{{figures/}}

\newsavebox\CBox

\input{KIT_colors.tex}

\begin{document}

\title{The ARDoCo Tool Landscape: REST API, TraceView, and TraceViz for Architecture Traceability}


\author{Jan Keim}
\orcid{0000-0002-8899-7081}
\email{jan.keim@kit.edu}
\author{Dominik Fuchß}
\orcid{0000-0001-6410-6769}
\email{dominik.fuchss@kit.edu}
\affiliation{%
  \institution{Karlsruhe Institute of Technology}
  \city{Karlsruhe}
  \country{Germany}
}

\author{Sophie Corallo}
\orcid{0000-0002-1531-2977}
\email{sophie.corallo@kit.edu}
\author{Tobias Hey}
\orcid{0000-0003-0381-1020}
\email{hey@kit.edu}
\affiliation{%
  \institution{Karlsruhe Institute of Technology}
  \city{Karlsruhe}
  \country{Germany}
}

\author{Julian Winter}
\orcid{0009-0001-4436-1081}
\email{julian.winter@student.kit.edu}
\author{Kevin Feichtinger}
\orcid{0000-0003-1182-5377}
\email{kevin.feichtinger@kit.edu}
\affiliation{%
  \institution{Karlsruhe Institute of Technology}
  \city{Karlsruhe}
  \country{Germany}
}


\begin{abstract}
\emph{Context and Problem.}
Software development produces interrelated artifacts like \SAD, \SAMs, and source code, whose relationships are essential for maintenance and consistency checking.
However, automatically recovering links between these artifacts (\TLR) remains difficult to deploy in practice.
\\\emph{Method and Aim.}
We present an accessible tool landscape for ARDoCo's \TLR approaches: the \textbf{ARDoCo REST API} exposes four \TLR pipelines (\SAD-\SAM, \SAM-Code, \SAD-Code, and \SAD-\SAM-Code) via HTTP endpoints with asynchronous execution and caching; \textbf{TraceView} is a browser-based frontend with a guided wizard and interactive multi-panel exploration of recovered links and inconsistencies; and \textbf{TraceViz}, which is a VS~Code extension that overlays trace links directly onto documentation in the IDE.
\\\emph{Results and Conclusion.}
All three components are publicly deployed and usable.
A preliminary study for TraceViz's in-IDE visualization confirmed that it improves developer comprehension during software understanding tasks.
The tool landscape makes state-of-the-art \TLR accessible to architects, developers, and tool integrators.
\\\emph{Video.}
We provide a screencast of our ARDoCo Tool Landscape and how it is used here: \url{https://youtu.be/IOTEPZQ3tVs}
\end{abstract}




\maketitle

\input{sections/introduction}
\input{sections/background}

\input{sections/tool}
\input{sections/evaluation}
\input{sections/conclusion}

\begin{acks}
This work was also supported by funding from the pilot program Core Informatics at KIT (KiKIT) of the Helmholtz Association (HGF), the Deutsche Forschungsgemeinschaft (DFG, German Research Foundation) – SFB 1608 – 501798263, the Topic Engineering Secure Systems of the HGF, and supported by KASTEL Security Research Labs, Karlsruhe.
Generative AI tools were used for copy editing.
\end{acks}

\bibliographystyle{ACM-Reference-Format}
\bibliography{ase26}


\end{document}

%% file: acronyms.tex
\begin{acronym}
    \acro{NLP}{natural language processing}
    \acro{IR}{information retrieval}
    \acro{LLM}{large language model}
    \acro{TLR}{traceability link recovery}
    \acro{RAG}{retrieval-augmented generation}
    \acro{SAD}{software architecture documentation}
    \acro{SAM}{software architecture model}
    \acro{TEAM}{Text Entity Absent from Model}
    \acro{MEAT}{Model Entity Absent from Text}
    \acro{SWATTR}{SoftWare Architecture Text Trace link Recovery}
    \acro{ArCoTL}{Architecture--Code Trace Links}
    \acro{LiSSA}{Linking Software System Artifacts}
    \acro{TransArC}{Transitive Architecture-to-Code}
\end{acronym}

\newcommand{\NLP}{\ac{NLP}\xspace}
\newcommand{\IR}{\ac{IR}\xspace}
\newcommand{\LLM}{\ac{LLM}\xspace}
\newcommand{\LLMs}{\acp{LLM}\xspace}
\newcommand{\TLR}{\ac{TLR}\xspace}
\newcommand{\RAG}{\ac{RAG}\xspace}
\newcommand{\SAD}{\ac{SAD}\xspace}
\newcommand{\SADs}{\acp{SAD}\xspace}
\newcommand{\SAM}{\ac{SAM}\xspace}
\newcommand{\SAMs}{\acp{SAM}\xspace}
\newcommand{\IDE}{IDE\xspace}
\newcommand{\REST}{REST\xspace}
\newcommand{\TEAM}{\ac{TEAM}\xspace}
\newcommand{\MEAT}{\ac{MEAT}\xspace}
\newcommand{\SWATTR}{\ac{SWATTR}\xspace}
\newcommand{\ArCoTL}{\ac{ArCoTL}\xspace}
\newcommand{\LiSSA}{\ac{LiSSA}\xspace}
\newcommand{\TransArC}{\ac{TransArC}\xspace}

%% file: KIT_colors.tex
\definecolor{kit-green}{RGB}{0, 150, 130}
\definecolor{kit-green100}{RGB}{0, 150, 130}
\definecolor{kit-green90}{rgb}{0.1, 0.6294, 0.5588}
\definecolor{kit-green80}{rgb}{0.2, 0.6706, 0.6078}
\definecolor{kit-green75}{rgb}{0.25, 0.6912, 0.6324}
\definecolor{kit-green70}{rgb}{0.3, 0.7118, 0.6569}
\definecolor{kit-green60}{rgb}{0.4, 0.7529, 0.7059}
\definecolor{kit-green50}{rgb}{0.5, 0.7941, 0.7549}
\definecolor{kit-green40}{rgb}{0.6, 0.8353, 0.8039}
\definecolor{kit-green30}{rgb}{0.7, 0.8765, 0.8529}
\definecolor{kit-green25}{rgb}{0.75, 0.8971, 0.8775}
\definecolor{kit-green20}{rgb}{0.8, 0.9176, 0.902}
\definecolor{kit-green15}{rgb}{0.85, 0.9382, 0.9265}
\definecolor{kit-green10}{rgb}{0.9, 0.9588, 0.951}
\definecolor{kit-green5}{rgb}{0.95, 0.9794, 0.9755}

\definecolor{kit-blue}{RGB}{70, 100, 170}
\definecolor{kit-blue100}{RGB}{70, 100, 170}
\definecolor{kit-blue90}{rgb}{0.3471, 0.4529, 0.7}
\definecolor{kit-blue80}{rgb}{0.4196, 0.5137, 0.7333}
\definecolor{kit-blue75}{rgb}{0.4559, 0.5441, 0.75}
\definecolor{kit-blue70}{rgb}{0.4922, 0.5745, 0.7667}
\definecolor{kit-blue60}{rgb}{0.5647, 0.6353, 0.8}
\definecolor{kit-blue50}{rgb}{0.6373, 0.6961, 0.8333}
\definecolor{kit-blue40}{rgb}{0.7098, 0.7569, 0.8667}
\definecolor{kit-blue30}{rgb}{0.7824, 0.8176, 0.9}
\definecolor{kit-blue25}{rgb}{0.8186, 0.848, 0.9167}
\definecolor{kit-blue20}{rgb}{0.8549, 0.8784, 0.9333}
\definecolor{kit-blue15}{rgb}{0.8912, 0.9088, 0.95}
\definecolor{kit-blue10}{rgb}{0.9275, 0.9392, 0.9667}
\definecolor{kit-blue5}{rgb}{0.9637, 0.9696, 0.9833}

\definecolor{kit-red}{RGB}{162, 34, 35}
\definecolor{kit-red100}{RGB}{162, 34, 35}
\definecolor{kit-red90}{rgb}{0.6718, 0.22, 0.2235}
\definecolor{kit-red80}{rgb}{0.7082, 0.3067, 0.3098}
\definecolor{kit-red75}{rgb}{0.7265, 0.35, 0.3529}
\definecolor{kit-red70}{rgb}{0.7447, 0.3933, 0.3961}
\definecolor{kit-red60}{rgb}{0.7812, 0.48, 0.4824}
\definecolor{kit-red50}{rgb}{0.8176, 0.5667, 0.5686}
\definecolor{kit-red40}{rgb}{0.8541, 0.6533, 0.6549}
\definecolor{kit-red30}{rgb}{0.8906, 0.74, 0.7412}
\definecolor{kit-red25}{rgb}{0.9088, 0.7833, 0.7843}
\definecolor{kit-red20}{rgb}{0.9271, 0.8267, 0.8275}
\definecolor{kit-red15}{rgb}{0.9453, 0.87, 0.8706}
\definecolor{kit-red10}{rgb}{0.9635, 0.9133, 0.9137}
\definecolor{kit-red5}{rgb}{0.9818, 0.9567, 0.9569}

\definecolor{kit-yellow}{RGB}{252, 229, 0}
\definecolor{kit-yellow100}{RGB}{252, 229, 0}
\definecolor{kit-yellow90}{rgb}{0.9894, 0.9082, 0.1}
\definecolor{kit-yellow80}{rgb}{0.9906, 0.9184, 0.2}
\definecolor{kit-yellow75}{rgb}{0.9912, 0.9235, 0.25}
\definecolor{kit-yellow70}{rgb}{0.9918, 0.9286, 0.3}
\definecolor{kit-yellow60}{rgb}{0.9929, 0.9388, 0.4}
\definecolor{kit-yellow50}{rgb}{0.9941, 0.949, 0.5}
\definecolor{kit-yellow40}{rgb}{0.9953, 0.9592, 0.6}
\definecolor{kit-yellow30}{rgb}{0.9965, 0.9694, 0.7}
\definecolor{kit-yellow25}{rgb}{0.9971, 0.9745, 0.75}
\definecolor{kit-yellow20}{rgb}{0.9976, 0.9796, 0.8}
\definecolor{kit-yellow15}{rgb}{0.9982, 0.9847, 0.85}
\definecolor{kit-yellow10}{rgb}{0.9988, 0.9898, 0.9}
\definecolor{kit-yellow5}{rgb}{0.9994, 0.9949, 0.95}

\definecolor{kit-orange}{RGB}{223, 155, 27}
\definecolor{kit-orange100}{RGB}{223, 155, 27}
\definecolor{kit-orange90}{rgb}{0.8871, 0.6471, 0.1953}
\definecolor{kit-orange80}{rgb}{0.8996, 0.6863, 0.2847}
\definecolor{kit-orange75}{rgb}{0.9059, 0.7059, 0.3294}
\definecolor{kit-orange70}{rgb}{0.9122, 0.7255, 0.3741}
\definecolor{kit-orange60}{rgb}{0.9247, 0.7647, 0.4635}
\definecolor{kit-orange50}{rgb}{0.9373, 0.8039, 0.5529}
\definecolor{kit-orange40}{rgb}{0.9498, 0.8431, 0.6424}
\definecolor{kit-orange30}{rgb}{0.9624, 0.8824, 0.7318}
\definecolor{kit-orange25}{rgb}{0.9686, 0.902, 0.7765}
\definecolor{kit-orange20}{rgb}{0.9749, 0.9216, 0.8212}
\definecolor{kit-orange15}{rgb}{0.9812, 0.9412, 0.8659}
\definecolor{kit-orange10}{rgb}{0.9875, 0.9608, 0.9106}
\definecolor{kit-orange5}{rgb}{0.9937, 0.9804, 0.9553}

\definecolor{kit-lightgreen}{RGB}{140, 182, 60}
\definecolor{kit-lightgreen100}{RGB}{140, 182, 60}
\definecolor{kit-lightgreen90}{rgb}{0.5941, 0.7424, 0.3118}
\definecolor{kit-lightgreen80}{rgb}{0.6392, 0.771, 0.3882}
\definecolor{kit-lightgreen75}{rgb}{0.6618, 0.7853, 0.4265}
\definecolor{kit-lightgreen70}{rgb}{0.6843, 0.7996, 0.4647}
\definecolor{kit-lightgreen60}{rgb}{0.7294, 0.8282, 0.5412}
\definecolor{kit-lightgreen50}{rgb}{0.7745, 0.8569, 0.6176}
\definecolor{kit-lightgreen40}{rgb}{0.8196, 0.8855, 0.6941}
\definecolor{kit-lightgreen30}{rgb}{0.8647, 0.9141, 0.7706}
\definecolor{kit-lightgreen25}{rgb}{0.8873, 0.9284, 0.8088}
\definecolor{kit-lightgreen20}{rgb}{0.9098, 0.9427, 0.8471}
\definecolor{kit-lightgreen15}{rgb}{0.9324, 0.9571, 0.8853}
\definecolor{kit-lightgreen10}{rgb}{0.9549, 0.9714, 0.9235}
\definecolor{kit-lightgreen5}{rgb}{0.9775, 0.9857, 0.9618}

\definecolor{kit-purple}{RGB}{163, 16, 124}
\definecolor{kit-purple100}{RGB}{163, 16, 124}
\definecolor{kit-purple90}{rgb}{0.6753, 0.1565, 0.5376}
\definecolor{kit-purple80}{rgb}{0.7114, 0.2502, 0.589}
\definecolor{kit-purple75}{rgb}{0.7294, 0.2971, 0.6147}
\definecolor{kit-purple70}{rgb}{0.7475, 0.3439, 0.6404}
\definecolor{kit-purple60}{rgb}{0.7835, 0.4376, 0.6918}
\definecolor{kit-purple50}{rgb}{0.8196, 0.5314, 0.7431}
\definecolor{kit-purple40}{rgb}{0.8557, 0.6251, 0.7945}
\definecolor{kit-purple30}{rgb}{0.8918, 0.7188, 0.8459}
\definecolor{kit-purple25}{rgb}{0.9098, 0.7657, 0.8716}
\definecolor{kit-purple20}{rgb}{0.9278, 0.8125, 0.8973}
\definecolor{kit-purple15}{rgb}{0.9459, 0.8594, 0.9229}
\definecolor{kit-purple10}{rgb}{0.9639, 0.9063, 0.9486}
\definecolor{kit-purple5}{rgb}{0.982, 0.9531, 0.9743}

\definecolor{kit-brown}{RGB}{167, 130, 46}
\definecolor{kit-brown100}{RGB}{167, 130, 46}
\definecolor{kit-brown90}{rgb}{0.6894, 0.5588, 0.2624}
\definecolor{kit-brown80}{rgb}{0.7239, 0.6078, 0.3443}
\definecolor{kit-brown75}{rgb}{0.7412, 0.6324, 0.3853}
\definecolor{kit-brown70}{rgb}{0.7584, 0.6569, 0.4263}
\definecolor{kit-brown60}{rgb}{0.7929, 0.7059, 0.5082}
\definecolor{kit-brown50}{rgb}{0.8275, 0.7549, 0.5902}
\definecolor{kit-brown40}{rgb}{0.862, 0.8039, 0.6722}
\definecolor{kit-brown30}{rgb}{0.8965, 0.8529, 0.7541}
\definecolor{kit-brown25}{rgb}{0.9137, 0.8775, 0.7951}
\definecolor{kit-brown20}{rgb}{0.931, 0.902, 0.8361}
\definecolor{kit-brown15}{rgb}{0.9482, 0.9265, 0.8771}
\definecolor{kit-brown10}{rgb}{0.9655, 0.951, 0.918}
\definecolor{kit-brown5}{rgb}{0.9827, 0.9755, 0.959}

\definecolor{kit-cyan}{RGB}{35, 161, 224}
\definecolor{kit-cyan100}{RGB}{35, 161, 224}
\definecolor{kit-cyan90}{rgb}{0.2235, 0.6682, 0.8906}
\definecolor{kit-cyan80}{rgb}{0.3098, 0.7051, 0.9027}
\definecolor{kit-cyan75}{rgb}{0.3529, 0.7235, 0.9088}
\definecolor{kit-cyan70}{rgb}{0.3961, 0.742, 0.9149}
\definecolor{kit-cyan60}{rgb}{0.4824, 0.7788, 0.9271}
\definecolor{kit-cyan50}{rgb}{0.5686, 0.8157, 0.9392}
\definecolor{kit-cyan40}{rgb}{0.6549, 0.8525, 0.9514}
\definecolor{kit-cyan30}{rgb}{0.7412, 0.8894, 0.9635}
\definecolor{kit-cyan25}{rgb}{0.7843, 0.9078, 0.9696}
\definecolor{kit-cyan20}{rgb}{0.8275, 0.9263, 0.9757}
\definecolor{kit-cyan15}{rgb}{0.8706, 0.9447, 0.9818}
\definecolor{kit-cyan10}{rgb}{0.9137, 0.9631, 0.9878}
\definecolor{kit-cyan5}{rgb}{0.9569, 0.9816, 0.9939}

\definecolor{kit-gray}{RGB}{0, 0, 0}
\definecolor{kit-gray100}{RGB}{0, 0, 0}
\definecolor{kit-gray90}{rgb}{0.1, 0.1, 0.1}
\definecolor{kit-gray80}{rgb}{0.2, 0.2, 0.2}
\definecolor{kit-gray75}{rgb}{0.25, 0.25, 0.25}
\definecolor{kit-gray70}{rgb}{0.3, 0.3, 0.3}
\definecolor{kit-gray60}{rgb}{0.4, 0.4, 0.4}
\definecolor{kit-gray50}{rgb}{0.5, 0.5, 0.5}
\definecolor{kit-gray40}{rgb}{0.6, 0.6, 0.6}
\definecolor{kit-gray30}{rgb}{0.7, 0.7, 0.7}
\definecolor{kit-gray25}{rgb}{0.75, 0.75, 0.75}
\definecolor{kit-gray20}{rgb}{0.8, 0.8, 0.8}
\definecolor{kit-gray15}{rgb}{0.85, 0.85, 0.85}
\definecolor{kit-gray10}{rgb}{0.9, 0.9, 0.9}
\definecolor{kit-gray5}{rgb}{0.95, 0.95, 0.95}

%% file: sections/introduction.tex
\section{Introduction}\label{sec:introduction}
\acresetall

\begin{figure}
\centering
\includegraphics{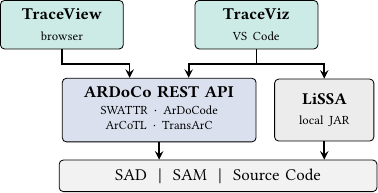}
\caption{ARDoCo tool landscape: \textbf{TraceView} and \textbf{TraceViz} serve different user needs. Both use the \textbf{\REST API}. \textbf{TraceViz} additionally supports \textbf{LiSSA}. Arrows indicate dependencies.}
\Description{Diagram of the ARDoCo tool landscape. Top row: TraceView (browser) and TraceViz (VS Code). Middle row: ARDoCo \REST API exposing SWATTR, ArDoCode, ArCoTL, and TransArC, and LiSSA (local JAR). Bottom row: input artifacts SAD, SAM, and Source Code. Arrows show TraceView connecting to the \REST API, TraceViz connecting to both the \REST API and LiSSA, and both the \REST API and LiSSA connecting to the artifact layer.}
\label{fig:landscape}
\end{figure}

Software systems are documented at multiple levels of abstraction: architects write natural-language \SADs to record principal design decisions, create \SAMs to formally specify components and their interactions, and developers implement the design in \emph{source code}.
Keeping these artifacts consistent and understandable, especially with regard to their mutual relations, is a prerequisite for effective maintenance, change impact analysis, and evolution~\cite{maeder12, cleland2012}.
Yet, manually maintaining trace links is tedious and error-prone, which makes automated \TLR essential in practice.
Automated \TLR is inherently difficult due to the abstraction gaps between natural language, models, and code.

The ARDoCo project\footnote{\url{https://ardoco.de}} develops and maintains a family of complementary, empirically validated \TLR approaches:
\acsu{SWATTR} (\acl{SWATTR})~\cite{keim_tracelink_2021} for \SAD-\SAM links;
an inconsistency detection approach based on \ac{SWATTR}~\cite{keim_detecting_2023};
\acsu{ArCoTL} (\acl{ArCoTL}) and \acsu{TransArC} (\acl{TransArC})~\cite{keim_recovering_2024} for \SAM-Code and transitive \SAD-Code links; and
\acsu{LiSSA} (\acl{LiSSA})~\cite{fuchss_lissa_2025} for \RAG-based \TLR across artifact types.
Recent work uses \LLMs to extract component names and identify named architecture entities~\cite{fuchss_enabling_2025,fuchss_whos_2026}.

Despite strong empirical results, these approaches are available only as Java libraries or command-line tools, requiring technical expertise to configure and run.
This barrier can prevent their adoption by architects and developers, who would benefit the most.

This paper closes that gap with a tool landscape that exposes the ARDoCo \TLR family through three complementary interfaces tailored to different user needs (\autoref{fig:landscape}):
\begin{enumerate}[leftmargin=*]
  \item \textbf{ARDoCo \REST API} for developers and tool integrators: a Spring Boot service exposing four \TLR pipelines via HTTP with asynchronous execution and result caching.
  \item \textbf{TraceView} for architects and non-developers: a zero-install browser frontend that shows all three artifact types side by side and exposes ARDoCo's inconsistency detection. TraceView is optimized for \SAD-\SAM-Code \TLR.
  \item \textbf{TraceViz} for developers: a VS~Code extension optimized for \SAD-Code \TLR that overlays trace links as gutter markers, enabling one-click navigation from documentation to linked code elements~\cite{winter2025traceviz}.
\end{enumerate}

The interfaces are publicly deployed and require little to no local installation, making \TLR accessible via browser, \IDE, or \REST API.

%% file: sections/background.tex
\section{Background: The ARDoCo TLR Approaches}\label{sec:background}

ARDoCo addresses \TLR between three categories of software artifacts: \SAD (natural language), \SAM (e.g.\ UML or Palladio component models), and source code.
The approaches below form the algorithmic foundation of the tool landscape presented in this paper.

\paragraph{SWATTR (\SAD-\SAM)}
\SWATTR~\cite{keim_tracelink_2021} is a framework that uses \NLP and heuristics in a multi-stage pipeline to recover trace links between sentences in a \SAD and components in a \SAM.
We later extended \SWATTR with detection of two types of artifact inconsistencies~\cite{keim_detecting_2023}: \TEAM, where a text entity has no counterpart in the model, and \MEAT, where a model entity has no counterpart in the text.

\paragraph{ArCoTL (\SAM-Code)}
\ArCoTL~\cite{keim_recovering_2024} recovers trace links between \SAMs and source code by transforming both into intermediate representations and combining text similarity with various heuristics.

\paragraph{TransArC / ArDoCode (\SAD-Code)}
\TransArC~\cite{keim_recovering_2024} composes the approaches \SWATTR and \ArCoTL transitively (\SAD $\to$ \SAM $\to$ Code) to recover direct trace links between \SADs and source code.
ArDoCode is a simpler variant that applies \SWATTR heuristics directly to code without requiring a \SAM.
The approach is easier to deploy but with lower performance.

\paragraph{LiSSA (generic)}
\LiSSA~\cite{fuchss_lissa_2025} is a generic \RAG-based \TLR approach: for each source artifact, it uses \IR to retrieve candidate targets, then queries an \LLM to confirm trace links, supporting multiple artifact type pairs.
Unlike the heuristic-based approaches, \LiSSA is not limited to the \SAD-\SAM-Code pipeline and has demonstrated strong performance on e.g. requirements-to-requirements \TLR~\cite{hey_requirements_2025}. 
However, it requires access to an \LLM.

All heuristic-based approaches were evaluated on a common benchmark~\cite{fuchss_establishing_2023} of five open-source Java projects: MediaStore, TeaStore, TEAMMATES, BigBlueButton, and JabRef.

%% file: sections/tool.tex
\begin{figure*}
  \centering
  \includegraphics[trim=0 1.75cm 0 0cm, width=\textwidth]{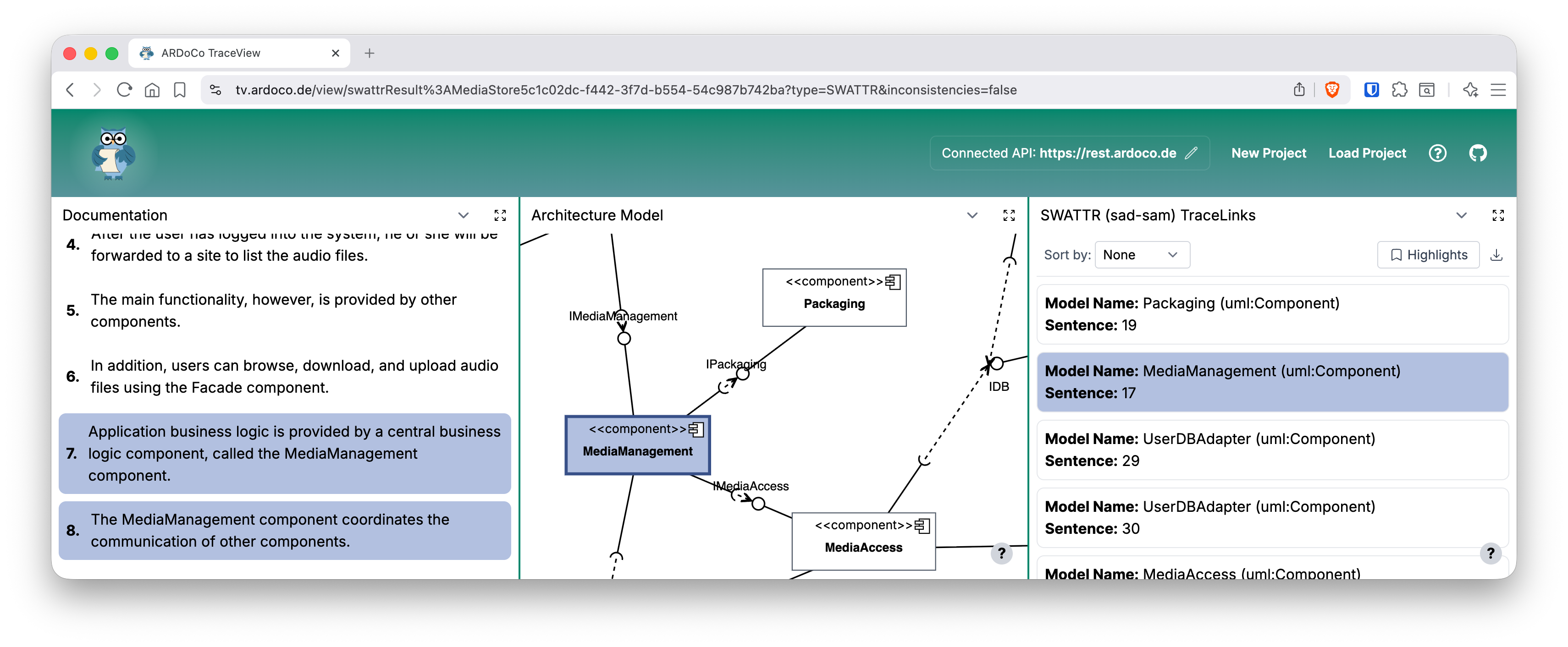}
  \caption{TraceView: \SAD documentation (left), \SAM (center), and recovered \SAD-\SAM trace links (right) shown side by side. Selecting a trace link highlights the linked \SAD sentence and SAM component across all panels.}
  \Description{Screenshot of ARDoCo TraceView in a browser showing three side-by-side panels: Documentation with numbered \SAD sentences, Architecture Model with a UML component diagram, and a \SWATTR trace link list. Sentence~7 and the MediaManagement component are highlighted in blue, indicating the selected trace link.}
  \label{fig:traceview}
\end{figure*}

\section{The ARDoCo Tool Landscape}\label{sec:tool}

The ARDoCo tooling addresses architecture traceability for different user needs through three complementary components: the \textbf{ARDoCo \REST API} as the shared computational backend, \textbf{TraceView} as an installation-free browser frontend optimized for \SAD-\SAM-Code \TLR with simultaneous multi-artifact visualization and inconsistency detection, and \textbf{TraceViz} as a VS~Code extension optimized for \SAD-Code \TLR with in-IDE navigation from documentation to code and support for our most recent \TLR approach \LiSSA.
All components are distributed as Docker images or extension packages and are publicly available.

\subsection{ARDoCo \REST API}

The ARDoCo \REST API provides programmatic access to ARDoCo's \TLR pipelines over HTTP, enabling any application to trigger \TLR analyses and retrieve structured results without bundling the Java libraries.
It is implemented in Java using Spring Boot and is publicly deployed\footnote{\url{https://rest.ardoco.de}}.
The API exposes the four ARDoCo \TLR pipelines as HTTP services organized in four controllers:
\begin{itemize}[leftmargin=*]
  \item \texttt{sad-sam}: \SAD-\SAM \TLR with \TEAM/\MEAT inconsistency detection (\SWATTR)
  \item \texttt{sam-code}: \SAM-Code \TLR (\ArCoTL)
  \item \texttt{sad-code}: \SAD-Code \TLR without \SAM (ArDoCode, baseline)
  \item \texttt{sad-sam-code}: transitive \SAD-Code \TLR via \SAM (\TransArC)
\end{itemize}
Each controller exposes multiple endpoints: \texttt{start} (fire-and-forget, returns a run ID), \texttt{start-and-wait} (synchronous), \texttt{get-result} (retrieve result by ID), and \texttt{wait-for-result} (long-poll for completion).
Pipeline runs are identified by a hash of the request content, enabling \emph{transparent caching}.
Meaning, results are stored in a Redis instance, and are reused for identical inputs without re-running the pipeline.
Cached results expire after 24~hours.
All endpoints return structured JSON and can be explored interactively via the integrated Swagger~UI.

\subsection{TraceView}

TraceView\footnote{\url{https://tv.ardoco.de}} is a browser-based, installation-free tool optimized for \SAD-\SAM-Code \TLR.
It displays all three artifact types simultaneously and exposes ARDoCo's inconsistency detection, making it the primary tool for architects exploring full-stack traceability.
TraceView is implemented with Next.js and React (TypeScript), and utilizes the ARDoCo \REST API.

Users create a new \TLR project through a four-step wizard:
(1)~\textit{Upload files}: upload your artifacts (e.g., a plain-text \SAD, an UML or Palladio Component \SAM, and a code model);
(2)~\textit{Project details}: enter a project name;
(3)~\textit{Configure}: select the \TLR pipeline based on the submitted artifact types (\SWATTR, ArDoCode, \ArCoTL, or \TransArC);
(4)~\textit{Summary}: review the configuration before start.

After submission, TraceView polls the \REST API and notifies the user when results are ready.
Results are shown in a configurable \emph{multi-panel view} with up to three resizable panels allowing the user to display the \SAD, the \SAM, and the source code as well as the detected trace links or inconsistencies side by side (cf. \autoref{fig:traceview}).
Selecting an element in any panel highlights its linked counterparts in the other panels, enabling cross-artifact navigation.

\subsection{TraceViz}

TraceViz\footnote{\url{https://github.com/ardoco/traceviz}} is a VS~Code extension optimized for \SAD-Code \TLR, placing trace links directly in the developer's editing context.
It visualizes links between natural language documentation and source code as inline gutter markers. Thus, enabling the navigation from any \SAD sentence to the linked code files and back without leaving the IDE.
Accessible from the VS~Code activity bar, it provides a split sidebar: the \emph{Traceability Approaches} panel for configuring and triggering \TLR, and the \emph{Trace History} panel listing past runs that can be re-activated with one click.

TraceViz supports three trace link sources: (i)~the \textbf{ARDoCo \REST API} (support for ArDoCode and \TransArC); (ii)~\textbf{CSV import} from any \TLR tool matching the format that the gold standards have \cite{fuchss_establishing_2023}, making TraceViz approach-agnostic; and (iii)~\textbf{LiSSA} via local JAR invocation.

\begin{figure*}
  \centering
  \includegraphics[trim=0 1.75cm 0 0cm, width=\textwidth]{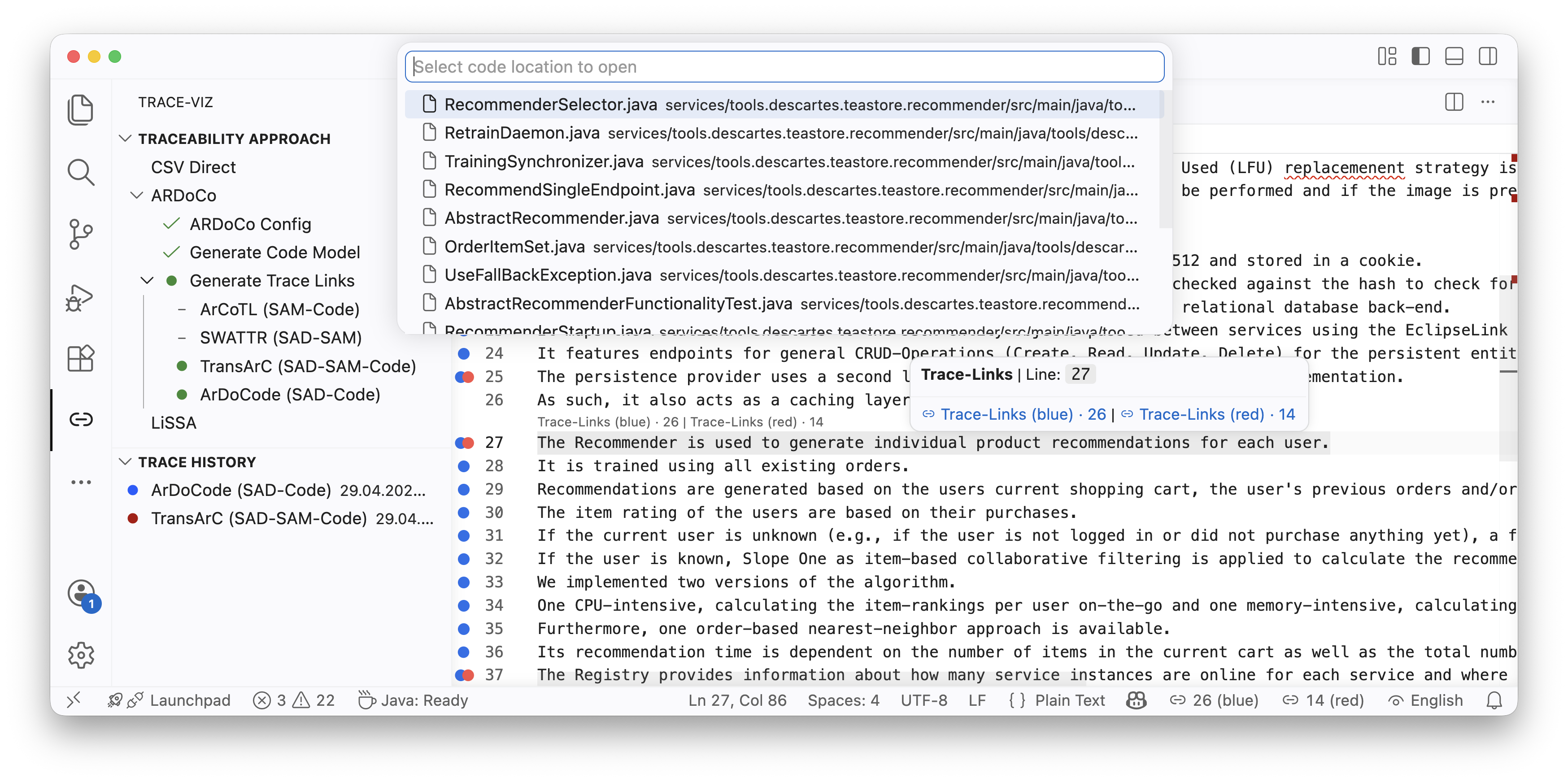}
  \caption{TraceViz in VS~Code: colored gutter dots mark \SAD-Code trace links. Hovering line~27 reveals per-set counts (blue:~26, red:~14) and a Quick~Pick listing linked code files for one-click navigation. The sidebar shows \TLR approaches and history.}
  \Description{Screenshot of VS Code with TraceViz active. The editor shows a documentation file with colored gutter dots (blue and red) on lines 24--37. A Quick Pick dialog lists Java source files as navigation targets. A hover tooltip on line~27 shows Trace-Links (blue) 26 and Trace-Links (red) 14. The left sidebar shows the TraceViz panel with a Traceability Approach tree (CSV Direct, ARDoCo sub-approaches \ArCoTL, \SWATTR, \TransArC, ArDoCode, and \LiSSA) and a Trace History section.}
  \label{fig:traceviz}
\end{figure*}

Once loaded, lines with links are marked with a colored \emph{gutter dot} (cf. \autoref{fig:traceviz}).
Hovering over these gutter dots reveals the link count, and clicking (or using the CodeLens annotation or status bar button) opens a Quick~Pick menu for navigating to linked artifacts.
Up to two trace link sets can be displayed simultaneously in selectable colors for side-by-side comparison.
A \emph{directory heuristic} replaces per-file markers with a single directory-level gutter dot when all files in a folder link to the same line, reducing visual noise.

%% file: sections/evaluation.tex
\section{Evaluation}\label{sec:evaluation}

\paragraph{Evaluations results for ARDoCo \TLR approaches}

The \TLR algorithms exposed by the \REST API have been evaluated in depth in their respective publications~\cite{keim_detecting_2023,keim_recovering_2024}.
On the ARDoCo benchmark~\cite{fuchss_establishing_2023} of five open-source Java projects (MediaStore, TeaStore, TEAMMATES, BigBlueButton, JabRef), \SWATTR achieves an average F1 of 0.81 for \SAD-\SAM, \ArCoTL achieves 0.98 for \SAM-Code, and \TransArC achieves 0.82 for \SAD-Code, significantly outperforming the best baseline (ArDoCode, F1\,=\,0.37)~\cite{keim_recovering_2024}.
Thus, according to the classification scheme of Hayes et al.~\cite{hayes_advancing_2006}, our best \TLR approaches can achieve excellent performance.
The inconsistency detection achieves F1\,=\,0.89 for \MEAT~\cite{keim_detecting_2023}.

\paragraph{Preliminary User Study of TraceViz}

To assess the usefulness of \IDE-integrated trace link visualization, Winter~\cite{winter2025traceviz} conducted a think-aloud user study of TraceViz ($n=7$; three doctoral candidates, three master's graduates, one industry developer with more than ten years of experience).
Participants completed two software comprehension tasks on the TEAMMATES project: one without visualization (raw CSV trace links) and one aided by TraceViz with \SAD-Code links.

Six of the seven participants found the visualizations \emph{helpful} in completing the tasks, and all seven agreed that the visualizations \emph{supported their comprehension process}.
Think-aloud recordings revealed that without visualization, participants relied on repetitive manual tree traversal, whereas TraceViz enabled direct jumps from documentation sentences to relevant source files.
Three participants stated that they would use this visualization in a real-world context. 
Four were undecided.
One participant highlighted the potential during onboarding scenarios, or during navigating large code bases where no guidance can be provided.

Even though there are limited participants in the current study, the results indicate that the participants perceived trace link visualization as a meaningful improvement to their efficiency and accuracy when navigating source code.

%% file: sections/conclusion.tex
\section{Conclusion}\label{sec:conclusion}

We presented the ARDoCo tool landscape, making a family of \TLR approaches available to software developers and architects: the \textbf{ARDoCo REST API}, \textbf{TraceView}, and \textbf{TraceViz}.
The \REST API exposes four \TLR pipelines (\SAD-\SAM via ARDoCo, \SAM-Code via \ArCoTL, \SAD-Code via ArDoCode, and transitive \SAD-\SAM-Code via \TransArC) via well-defined HTTP endpoints with asynchronous execution and Redis-backed caching.
TraceView provides a browser-based wizard for artifact upload and configuration, as well as an interactive multi-panel result view.
TraceViz integrates trace links directly into Visual Studio Code, enabling immediate navigation from documentation lines to linked code elements in the \IDE.
The preliminary think-aloud user study of TraceViz ($n=7$) found that visualization consistently improved developer comprehension: six of seven participants found the visualization \textit{helpful} when completing the task, and all seven reported that it supported their search or comprehension process~\cite{winter2025traceviz}.

In future work 
we plan to integrate the \LLM-based \TLR approach \LiSSA and the \LLM-based architecture component name extraction ExArch~\cite{fuchss_whos_2026} into the \REST API, making them accessible through both TraceView and TraceViz.
Further, we aim at extending support for additional artifact types (e.g., requirements), improving visualization for large projects with many trace links, and conducting larger, controlled user studies across all ARDoCo tools.

%% file: ase26.bib
@inproceedings{keim_tracelink_2021,
  author            = {Keim, Jan and Schulz, Sophie and Fuch{\ss}, Dominik and Kocher, Claudius and Speit, Janek and Koziolek, Anne},
  editor            = {Biffl, Stefan and Navarro, Elena and L{\"o}we, Welf and Sirjani, Marjan and Mirandola, Raffaela and Weyns, Danny},
 title="Trace Link Recovery for Software Architecture Documentation",
booktitle="Software Architecture",
year="2021",
publisher="Springer International Publishing",
isbn="978-3-030-86044-8"
}

@inproceedings{fuchss_establishing_2023,
  title             = {{Establishing} a {Benchmark} {Dataset} for {Traceability} {Link} {Recovery} {Between} {Software} {Architecture} {Documentation} and {Models}},
  author            = {Fuch{\ss}, Dominik and Corallo, Sophie and Keim, Jan and Speit, Janek and Koziolek, Anne},
  editor="Batista, Thais
and Bure{\v{s}}, Tom{\'a}{\v{s}}
and Raibulet, Claudia
and Muccini, Henry",
booktitle="Software Architecture. ECSA 2022 Tracks and Workshops",
year="2023",
publisher="Springer International Publishing",
address="Cham",
pages="455--464",
isbn="978-3-031-36889-9"
}

@inproceedings{keim_detecting_2023,
  title             = {{Detecting} {Inconsistencies} in {Software} {Architecture} {Documentation} {Using} {Traceability} {Link} {Recovery}},
  author            = {Keim, Jan and Corallo, Sophie and Fuch{\ss}, Dominik and Koziolek, Anne},
  abbr              = {ICSA},
  booktitle         = {IEEE 20th International Conference on Software Architecture (ICSA)},
  doi               = {10.1109/icsa56044.2023.00021},
  google_scholar_id = {WF5omc3nYNoC},
  number            = {},
  pages             = {141--152},
  preprint          = {2023/icsa-23-inconsistencies.pdf},
  publisher         = {IEEE},
  redirect          = {https://publikationen.bibliothek.kit.edu/1000158208/150680132},
  volume            = {},
  year              = {2023},
  month             = {3}
}

@inproceedings{keim_recovering_2024,
  title             = {{Recovering} {Trace} {Links} {Between} {Software} {Documentation} {And} {Code}},
  author            = {Keim, Jan and Corallo, Sophie and Fuch{\ss}, Dominik and Hey, Tobias and Telge, Tobias and Koziolek, Anne},
  address           = {New York, NY, USA},
  booktitle         = {Proceedings of the IEEE/ACM 46th International Conference on Software Engineering},
  doi               = {10.1145/3597503.3639130},
  google_scholar_id = {UebtZRa9Y70C},
  isbn              = {9798400702174},
  journal           = {ICSE},
  location          = {Lisbon, Portugal},
  month             = {4},
  numpages          = {13},
  pages             = {1--13},
  preprint          = {2024/icse-24-tlr.pdf},
  publisher         = {ACM},
  selected          = {true},
  year              = {2024}
}

@inproceedings{fuchss_enabling_2025,
  title             = {{Enabling} {Architecture} {Traceability} by {LLM}-based {Architecture} {Component} {Name} {Extraction}},
  author            = {Fuch{\ss}, Dominik and Liu, Haoyu and Hey, Tobias and Keim, Jan and Koziolek, Anne},
  abbr              = {ICSA},
  booktitle         = {IEEE 22nd International Conference on Software Architecture (ICSA)},
  doi               = {10.1109/icsa65012.2025.00011},
  eventdate         = {2025-03-31/2025-04-04},
  eventtitle        = {22nd IEEE International Conference on Software Architecture},
  eventtitleaddon   = {ICSA 2025},
  google_scholar_id = {7PzlFSSx8tAC},
  language          = {english},
  month             = {3},
  pages             = {1--12},
  publisher         = {IEEE},
  redirect          = {https://publikationen.bibliothek.kit.edu/1000179830/157561351},
  selected          = {true},
  venue             = {Odense, Denmark},
  year              = {2025}
}

@inproceedings{fuchss_lissa_2025,
  title             = {{LiSSA}: {Toward} {Generic} {Traceability} {Link} {Recovery} {Through} {Retrieval}-{Augmented} {Generation}},
  author            = {Fuch{\ss}, Dominik and Hey, Tobias and Keim, Jan and Liu, Haoyu and Ewald, Niklas and Thirolf, Tobias and Koziolek, Anne},
  abbr              = {ICSE},
  booktitle         = {IEEE/ACM 47th International Conference on Software Engineering (ICSE)},
  doi               = {10.1109/icse55347.2025.00186},
  google_scholar_id = {8k81kl-MbHgC},
  location          = {Ottawa, Canada},
  month             = {4},
  pages             = {1396--1408},
  preprint          = {2025/icse-25.pdf},
  publisher         = {IEEE},
  redirect          = {https://publikationen.bibliothek.kit.edu/1000179816/157462526},
  selected          = {true},
  year              = {2025}
}

@inproceedings{hey_requirements_2025,
  title             = {{Requirements} {Traceability} {Link} {Recovery} via {Retrieval}-{Augmented} {Generation}},
  author            = {Hey, Tobias and Fuch{\ss}, Dominik and Keim, Jan and Koziolek, Anne},
  abbr              = {REFSQ},
  address           = {Cham},
  booktitle         = {Lecture Notes in Computer Science},
  doi               = {10.1007/978-3-031-88531-0\_27},
  editor            = {Hess, Anne and Susi, Angelo},
  google_scholar_id = {MXK_kJrjxJIC},
  isbn              = {978-3031885303},
  issn              = {0302-9743},
  journal           = {Requirements Engineering: Foundation for Software Quality},
  month             = {04},
  preprint          = {2025/refsq-25.pdf},
  publisher         = {Springer Nature Switzerland},
  redirect          = {https://publikationen.bibliothek.kit.edu/1000178589/156854596},
  year              = {2025}
}

@article{fuchss_whos_2026,
  author            = {Fuch{\ss}, Dominik and Liu, Haoyu and Corallo, Sophie and Hey, Tobias and Keim, Jan and von Geisau, Johannes and Koziolek, Anne},
  title             = {Who's Who? LLM-assisted Software Traceability with Architecture Entity Recognition},
  year              = {2026},
  publisher         = {Association for Computing Machinery},
  address           = {New York, NY, USA},
  issn              = {1556-4665},
  url               = {https://doi.org/10.1145/3807453},
  doi               = {10.1145/3807453},
  journal           = {ACM Trans. Auton. Adapt. Syst.},
  month             = apr,
  google_scholar_id = {mB3voiENLucC},
  abbr              = {TAAS}
}

@book{cleland2012,
  title     = {Software and Systems Traceability},
  editor    = {Jane Cleland-Huang and Orlena Gotel and Andrea Zisman},
  year      = {2012},
  publisher = {Springer London},
  address   = {London},
  doi       = {10.1007/978-1-4471-2239-5},
  url       = {https://doi.org/10.1007/978-1-4471-2239-5}
}

@inproceedings{maeder12,
  author    = {Mäder, Patrick and Egyed, Alexander},
  booktitle = {2012 28th IEEE International Conference on Software Maintenance},
  title     = {Assessing the effect of requirements traceability for software maintenance},
  year      = {2012},
  volume    = {},
  number    = {},
  doi       = {10.1109/ICSM.2012.6405269}
}

@mastersthesis{winter2025traceviz,
    author       = {Winter, Julian Robin},
    year         = {2025},
    title        = {Guided Exploration and Visualization of Trace Links in Visual Studio Code},
    doi          = {10.5445/IR/1000192928},
    publisher    = {{Karlsruher Institut für Technologie (KIT)}},
    pagetotal    = {62},
    type    = {Bachelor's Thesis},
    school       = {Karlsruher Institut für Technologie (KIT)},
    language     = {english}
}

@article{hayes_advancing_2006,
  title = {Advancing {{Candidate Link Generation}} for {{Requirements Tracing}}: {{The Study}} of {{Methods}}},
  shorttitle = {Advancing {{Candidate Link Generation}} for {{Requirements Tracing}}},
  author = {Hayes, Jane Huffman and Dekhtyar, Alex and Sundaram, Senthil Karthikeyan},
  year = {2006},
  month = jan,
  journal = {IEEE TSE},
  volume = {32},
  number = {1},
  pages = {4--19},
  issn = {0098-5589},
  doi = {10.1109/TSE.2006.3},
  urldate = {2018-11-06}
}
